# Application of Artificial Neural Network in Jitter Analysis of Dispersion-Managed Communication System

F.P. Zen[1], B.E. Gunara[1], W. Hidayat[1], Z.A. Thalib[2], H. Zainuddin[2], J. Aminuddin[1,3] *

[1]Theoretical Physics Laboratory, Department of Physics,
Institute of Technology Bandung, Bandung, Indonesia.

[2]Department of Physics, Universiti Putra Malaysia, Selangor Darul Ehsan, Malaysia

[3]Department of Physics, Hasanuddin University, Makasar, Indonesia.

## ABSTRACT

Artificial Neural Network (ANN) is used as numerical methode in solving modified Nonlinear Schrödinger (NLS) equation with Dispersion Managed System (DMS) for jitter analysis. We take the optical axis $z$ and the time $t$ as input, and then some relevant values such as the change of position and the center frequency of the pulse, and further the mean square time of incoming pulse which are needed for jitter analysis. It shows that ANN yields numerical solutions which are adaptive with respect to the numerical errors and also verifies the previous numerical results using conventional numerical method. Our result indicates that DMS can minimize the timing jitter induced by some amplifiers.

---

* Email: fpzen@fi.itb.ac.id, bobby@fi.itb.ac.id, hisham@fsas.upm.edu.my

# I. Introduction

In data transmission using soliton communication system, some amplifiers are needed to overcome the loss fiber. Latter a new problem emerges that they also induce a number of noise which is significant enough to signal. The pulse accompanying noise which causes the uncertainty of the arrival time of soliton randomly is called jitter. It has statistically been shown by Gordon and Haus that jitter which emerges spontaneously from the effect of the noise from amplifiers is closely like Gaussian distribution with the variance is proportional to the square of data transmission travelled distance [1].

Unfortunately Gordon-Haus model is disagree with the experiment since there is a significant deviation in Gaussian-like distribution [2]. It might be that the additional physical effects such as soliton interaction, accoustic effect, and polarization mode dispersion also contribute to this deviation. For example, it was shown in [2] that soliton interaction plays a dominant role in the fast velocity data transmision.

On the other hand since the existence of jitter faces us a problem in detecting soliton, so we have to find a way to minimize the jitter. By exploiting the data transmision system where the nonlinear effect is compensated by a periodic dispersion (called dispersion-managed communication systems) it turns out that the jitter can be minimized. The enhancement power in Dispersion Managed System (DMS) causes the jitter reduction [3],[4].

The purpose of this paper is to verify the previous results in references [4,5] using Artificial Neural Network (ANN) as the numerical tool. This methode has been tested to solve many problems in partial differential equation and ordinary differential equation [6] because it has universal approximation property and further, is adaptive with numerical error [5]. ANN will be used in a different way, that is the trial function is ruled out. Since the object is highly complex and contains nonlinear terms, hence the numerical solution fully contains ANN terms involving some boundary and initial conditions on the definiton of training error. This method has been succeeded to find soliton solutions from NLS equation [7]. An extension result in DMS using ANN to find breathers soliton solutions has been done in reference [8].

This paper is devoted to study further the application of ANN in order to observe the existence of jitter which is induced by the noise of amplifier. It consist of four parts, that is Introduction in the first part, the second part is the ANN alogarithm, the discussion of our numerical results and its interpretation in the third part, and finally the conclusion.

**II. ANN Method and Numerical Solution of Periodic Dispersion NLS Equation**

In principle the application of ANN for time jitter analysis is how to find a numerical solution of nonlinear differential equation, that is NLS which has modified as follows [3]:

$$i\frac{\partial \psi}{\partial Z} + \frac{\sigma(Z)}{2}\frac{\partial^2 \psi}{\partial T^2} + |\psi|^2 \psi = iS(Z,T), \qquad (1)$$

where $S(Z,T)<<1$ is the white noise process and $\psi(Z,T)$ is normalized with respect to $|E_0|^2=(\lambda_0 A_{\text{eff}})/(2\pi n_2 Z_0)$. The time variabel $T$ and the travelled distance are also normalized with respect to $T_0/1.76$, where $T_0$ is the Full Width at Half Maximum (FWHM) and the dispersion length $Z_0 = (2\pi c)/(\lambda_0^2 \overline{D})(T_0/1.76)^2$ respectively, while $\overline{D}$ is the dispersion mean value of the dispersion pattern which is setted to the optical fiber [9]. Nonlinear diffraction index coefficient is denoted by $n_2$, while $\lambda_0$ is the wavelength carrier, dan $c$ is the speed of light. The dispersion pattern $\sigma(Z)$ which is a periodic value comes from anomalous dispersion and normal dispersion effects. In our simulation $\sigma(Z)$ will be taken from [9].

The numerical solution of equation (1) which has previously been proposed using ANN is [6,7]

$$\psi^q(x_i) = \psi^q{}_R(x_i) + i\psi^q{}_I(x_i) \qquad (2)$$

where $\psi^q{}_R = fo(\sum_h^H W^q_{ho} y^q_h)$ for o=1 and $\psi^q{}_I = fo(\sum_h^H W^q_{ho} y^q_h)$ for o=2. Furthermore $y^q{}_h(z,t,1) = f_H(\sum_{i=1}^{3} V^q_{ih} x^q_i)$ is the output value from the neuron at hidden layer. The activation functions are chosen to be $f_H(u) = \tanh(u) = \dfrac{1-e^{-2u}}{1+e^{-2u}}$ and $fo(u) = u$. The ANN design can be seen in Figure 1.

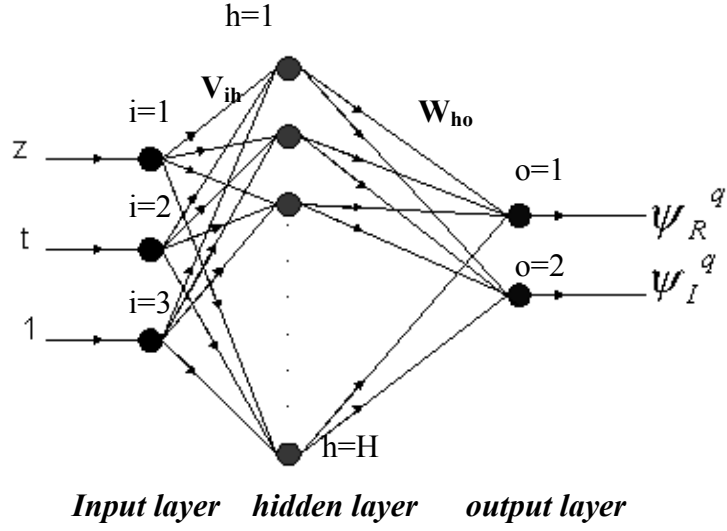

*Input layer   hidden layer    output layer*

Figure 1. ANN Structure

The purpose of training on every neuron is to find weight for which the ANN solution satisfies NLS equation together with its given boundary condition. Renewal of the weight is formulated by

$$V_{ih}^{q+1} = V_{ih}^{q} - \eta \frac{\partial E^q}{\partial V_{ih}^{q}}, \tag{3a}$$

$$W_{ho}^{q+1} = W_{ho}^{q} - \eta \frac{\partial E^q}{\partial W_{ho}^{q}}. \tag{3b}$$

Symbol $\eta$ in equation (3) is the speed of convergence of error value which goes to the lowest minimum (called learning rate). The value is chosen to be $0<\eta<1$ which depends on the requirement to have a fast calculation. The definition of $E^q$ as the error value on every training to $q$ is given by :

$$E^q = E_1^q + E_2^q + E_3^q = |F^q|^2 + |B_1^q|^2 + |B_2^q|^2. \tag{4}$$

A complete explaination of equation (4) is the following :

**(i).** $E_1^q = |F^q|^2$ is defined in equation (1),

$$F^q = i\frac{\partial \psi^q}{\partial Z} + \frac{\sigma(Z)}{2}\frac{\partial^2 \psi^q}{\partial T^2} + |\psi^q|^2 \psi^q - iS = 0 \tag{5}$$

where $\psi^q$ is written in equation (2),

$$\psi^q(x_i) = \psi^q_R(x_i) + i\psi^q_I(x_i) = fo(\sum_h^H W^q_{h1} y^q_h) + i\, fo(\sum_h^H W^q_{h2} y^q_h) = \sum_h^H W^q_{h1} y^q_h + i\sum_h^H W^q_{h2} y^q_h$$

$$= \sum_h^H W^q_{h1}\left[\tanh(\sum_{i=1}^3 V^q_{ih} x^q_i)\right] + i\sum_h^H W^q_{h2}\left[\tanh(\sum_{i=1}^3 V^q_{ih} x^q_i)\right]. \quad (6)$$

The value $E_1^q = |F^q|^2$ can be computed by inserting equation (6) to equation (5).

**(ii).** $E_2^q = |B_1^q|^2$ corresponds to the initial wave. Intensity of ANN result on $Z=0$, $T$ should be close to intensity of a given ansatz, so that the difference is also close to zero,

$$|B_1^q|^2 = \left(\psi_R^q(0,t)\right)^2 + \left(\psi_I^q(0,t)\right)^2 - \psi_0 \psi_0^* \quad (7)$$

where the first and the second terms are a training result and the third term is the given ansatz. For example

$$\psi(Z,T) = A\sqrt{\alpha}\,\mathrm{sec}\,h(\alpha[T-C])e^{i\left(\Omega[T-C]+\beta[T-C]^2+\frac{\phi}{2}\right)}, \quad (8)$$

with $\psi_0 = \psi(Z=0,T)$. In the above equation, the increasing power factor is denoted by $A$, while $\alpha$, $\beta$, $C$, $\Omega$, dan $\phi$ are $Z$ dependent parameter which denote amplitude, quadratic chirp, center-position, center-frequency, and dynamical phase, respectively.

**(iii).** $E_3^q = |B_2^q|^2$ is defined as intensity on every $Z$ for ($|T| \to \infty$) and its derivative with respect to $T$ should be close to zero.

$$|B_2^q|^2 = \left(\psi_R^q(Z,T\to\pm\infty)\right)^2 + \left(\psi_I^q(Z,T\to\pm\infty)\right)^2 + \left(\frac{\partial\psi_R^q(Z,T\to\pm\infty)}{\partial T}\right)^2 + \left(\frac{\partial\psi_I^q(Z,T\to\pm\infty)}{\partial T}\right)^2$$

(9)

Computer simulation can directly be done to get solution of equation (1) using ANN methode. Computation process is proceeded by computing equation (6) and then equation (4). If the result is closely to zero with our demanding tolerance then take the value of $\psi$ and write at $q^{th}$ traning. If this criteria is not satisfied, then the weight must be renewed by computing equation (3).

### III. Timing Jitter Analysis
### A. Amplifier Induced Jitter

Consider now equation (1) with non zero $S(Z,T)$ which is given in [3], that is

$$S(Z,T) = \sum_{n=0}^{N} \delta(Z+nP)\sqrt{\alpha^{(n)}} \operatorname{sec}h(\alpha^{(n)}[T-C^{(n)}])$$
$$\times \exp(i\varphi^{(n)})\{(S_1^{(n)}+iS_2^{(n)})\tanh(\alpha^{(n)}[T-C^{(n)}])+ \quad (10)$$
$$(S_3^{(n)}+iS_4^{(n)})\}$$

where $\alpha^{(n)}$, $\beta^{(n)}$, $\varphi^{(n)}$ dan $C^{(n)}$ are parameters on every $n^{th}$ periodic dispersion. The change of the center frequency can then be computed as

$$\Omega^{(n+1)} = \Omega^{(n)} - 2\frac{S_1^{(n)}\alpha^{(n)}}{3A}, \quad (11)$$

while the center-position is changed like

$$C^{(n+1)} = C^{(n)} - \frac{2S_1^{(n)}\alpha^{(n)}}{3A}P + \frac{S_2^{(n)}}{A\alpha^{(n)}}. \quad (12)$$

The mean square of the time difference of the incoming soliton has form

$$\langle \delta T^2 \rangle \approx \frac{4\alpha^2 \langle S_1^{(n)} \rangle}{9A^2} \frac{L^3}{3P}. \quad (13)$$

Clearly from equation (13), the timing jitter can be minimized by increasing energy denoted by *A*. This enhancement is caused by the DMS system. Timing jitter fluctuation tends smaller than ordinary soliton system which shown in Figure 2. Our ANN simulation result coincides with the result in [4], and the variable values dan are taken to be the same.

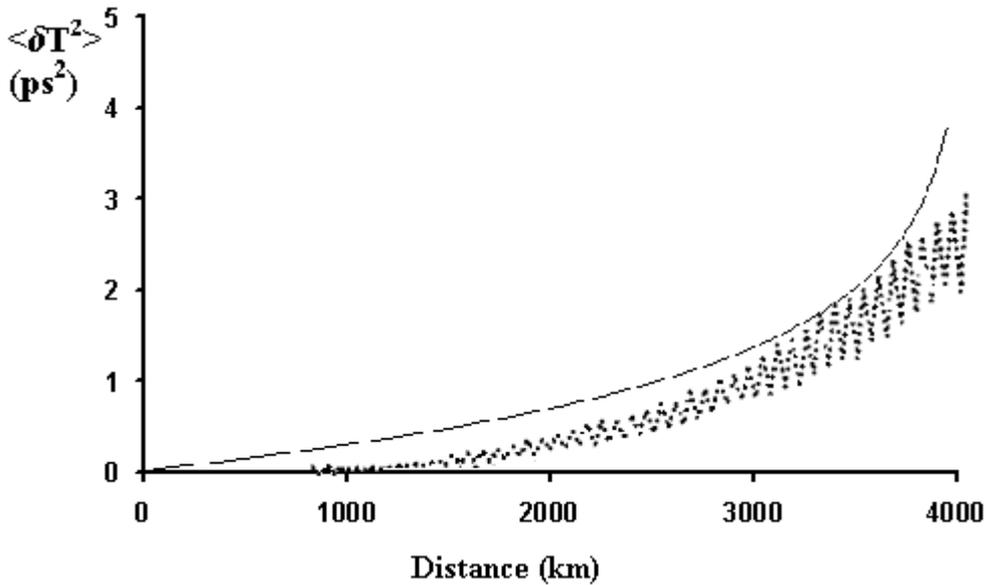

Figure 2. The plot of ANN method at $q=1400$ and Mean Squared Error (MSE) has order 0,01. Dashed-curve is jitter phenomenon on system without dispersion-managed. Doted-curve shows *jitter* phenomenon on DMS.

During ANN training process the change of Mean-Squared Error (MSE) is plotted in Figure 3. The process is stopped when $q=1400$ where MSE has order 0,01.

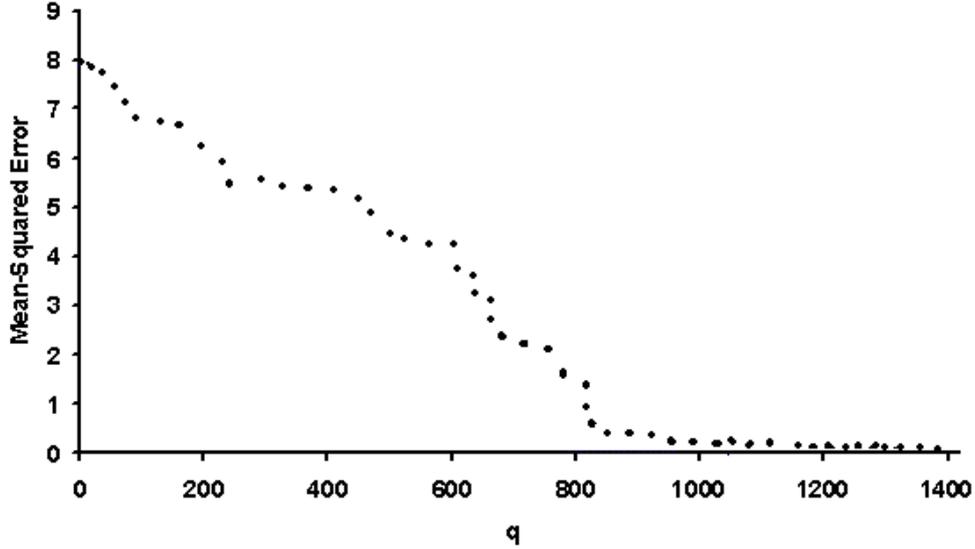

Figure 3. Mean-Squared Error (MSE) during ANN training process

**B. Influence of Amplifier Distance in Jitter Phenomenon**

During transmision the distance between amplifier for every distance $Z=Z_r$ ($Z_r$ is the distance between amplifier) can also influence the change of the pulse frequency denoted by $\delta\Omega$. The mean of $\delta\Omega$ is [4]

$$\langle\delta\Omega\rangle = \frac{N_{sp}(G-1)}{\sqrt{2\pi}\,N_0 A^2}\left(4p_r^2 + \frac{C_r^2}{p_r^2}\right), \tag{14}$$

where $N_{sp}$ is emission spontan factor, $G$ is amplifier gain, $N_0$ denotes a number of photon per unit energy, $p_r$ lis pulse width at $Z=Z_r$, and $C_r$ is chirp at $Z=Z_r$. In our simulation we choose $N_{sp}=1$, $Z_r=72$ km, the value of $C_r$ and $p_r$ can directly computed from NLS equation (1). ANN simulation is shown in Figure 4, where the change of the pulse frquency becomes minimum if the amplifier is located at every end of *dispersion map*. All parameters are taken to be the same with [5], and we regain the previous results in [5]. Figure 5 shows the fluctuation of Mean Squared Error (MSE) until it is stopped when the computaion process reach 0,01 that is when q=450.

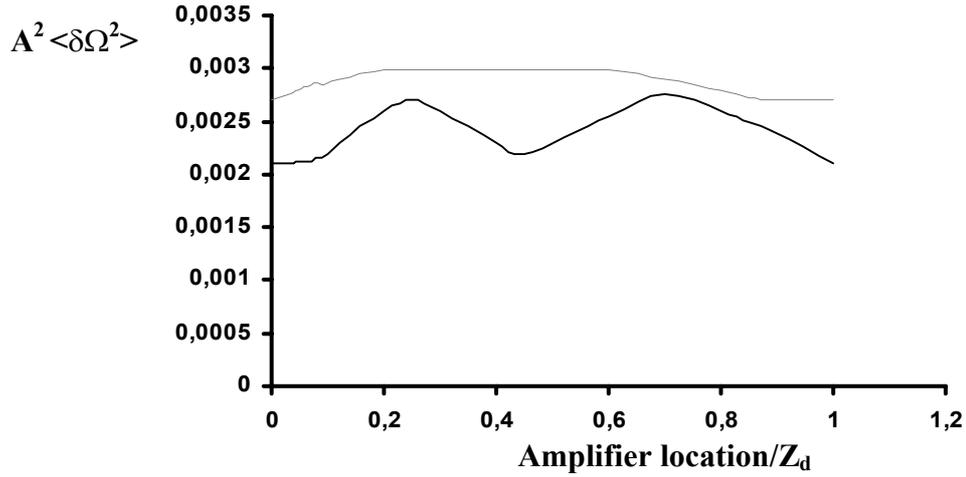

Figure 4. The change of mean frequency times $A^2$ as a function of amplifiers location in compensated dispersion scale $Z_d=Z_r=72$km. ANN is trained up to iteration i q=450 and MSE orde is 0,01.

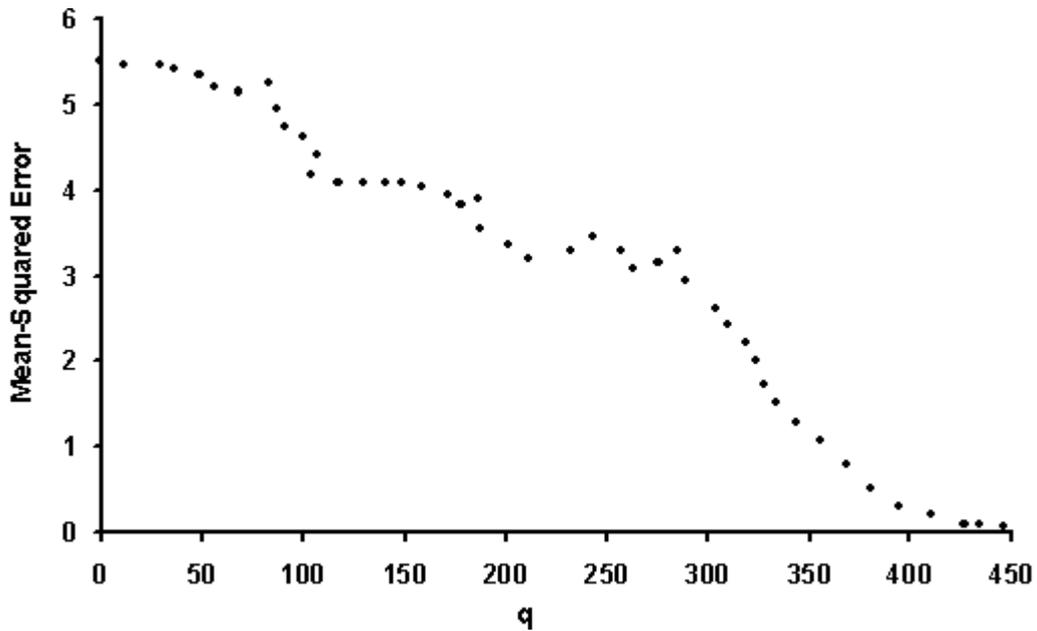

Figure 5. Mean-Squared Error (MSE) during ANN training process

## IV. Conclusions

It has been shown that ANN as universal approximation method gives a good description of numerical solution of the dispersion managed NLS equation. The ANN simulation verifies the previous result using conventional numeical methode [4,5]. For every input $z$ as the optical axis and the $t$, it results numerically the mean square of the time difference of the incoming $<\delta T^2>$ which also indicates the jitter phenomenon. Our numerical analysis shows that the jitter can be reduced using dispersion managed systems, see Figure 2. We also have discussed the influence of amplifiers location with respect to the change of pulse frequency which is shown in Figure 4, where the change of pulse frequency becomes minimum if the amplifiers are placed at every end of dispersion map. It also recommends to reduce jitter in the way of minimizing the change of pulse frequency.